\documentclass{article}
\usepackage{graphicx}
\usepackage{array}
\usepackage{epstopdf}
\usepackage{tabularx}
\usepackage{caption}
\begin{document}
\def\thepage{}

\title{TOPO: Improving remote homologue recognition via identifying common protein structure framework} 

\author{Jianwei Zhu$^1$, Haicang Zhang$^1$, Chao Wang$^{1}$, Bin Ling$^1$, \\
	Wei-Mou Zheng$^{2,*}$, Dongbo Bu$^1,$\footnote{Correspondence should be addressed to Wei-Mou Zheng (zheng@itp.ac.cn) and Dongbo Bu (dbu@ict.ac.cn)}\\
1 Key lab of intelligent information processing, \\
	Institute of Computing Technology, \\
	Chinese Academy of Sciences, Beijing, China\\
2 Institute of Theoretical Physics, \\
	Chinese Academy of Sciences, Beijing, China
}


\maketitle

\baselineskip=13.5pt

\begin{abstract}

\begin{center}
(Extended Abstract)
\end{center}

\end{abstract}

\baselineskip=13.5pt

\section{Motivation}

Protein structure prediction plays an important role in the fields of bioinformatics and biology. Traditional
protein structure prediction approaches include template-based modeling (TBM, including homology modeling,
and threading), and free modeling (FM). In particular, a threading algorithm takes a query protein sequence as input, recognizes
the most likely fold, and finally reports the alignments of the query sequence to structure-known
templates as output. The existing threading approaches mainly utilizes the information of protein
sequence profile, solvent accessibility, contact probability, etc. The threading strategy has been shown to be successful in structure prediction of a great amount of proteins; however, the existing threading approaches show poorly performance for remote homology
proteins. How to improve the fold recognition for remote homology proteins remains a challenge 
to protein structure prediction.

The sequences of proteins in remote homology generally show relatively weak signal of structure.
However, this does not mean that there is no sequence conservation hints for structure. The success of
multiple-templates strategy implies the existence of common frameworks, i.e. some regions of proteins
are conservative both in the structure and sequence. Such common frameworks should be responsible
to the structural stability and then conservative in the evolution. 

Based on this we proposed a novel threading approach in three steps. First, for each template, the common structural frameworks shared by its homologous proteins were calculated. Second, unlike in traditional threading methods where the alignment is made against the whole template, we aligned the query protein sequence against a common framework first. This strategy avoids the drawback
of the traditional threading approach, i.e. the alignment of variable regions beyond conserved motifs is
prone to bringing in error. Third, the final alignments were generated via aligning query sequence against candidate full-length templates in the family. Briefly speaking, we run TreeThreader\cite{TreeThreader} to build alignments of query against the new template database, and ranked alignments by E-value for model generation. Finally, we generated models by MODELLER based on 
candidate alignments. The generated models are ranked according to dDFIRE\cite{yang2008specific} energy function.

\section{Methods}

For each template with known structure, all of its remote homology proteins are first identified based
on structure alignment. Then, a linear programming was designed to identify the common framework shared
by these remote homology proteins.

\begin{figure*}
\centering
\includegraphics[width=5in]{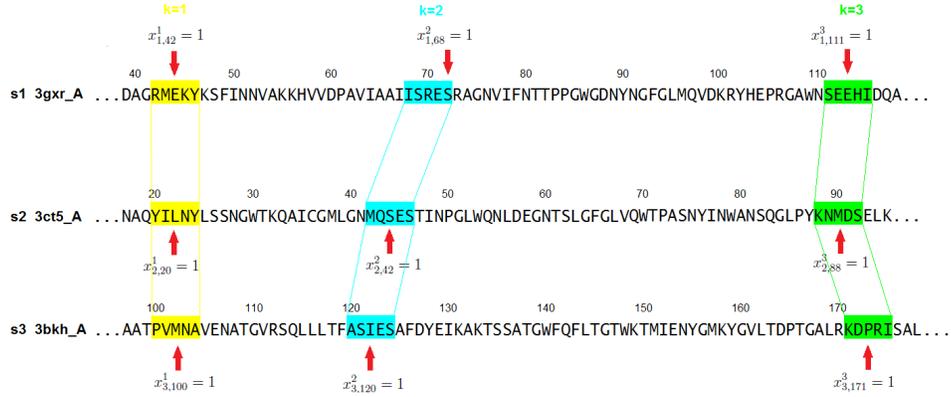}
\caption{Common frameworks shared by protein  {\tt 3gxr\_A} and its homologous proteins. The common framework consists of three dispersed segments (in yellow, cyan, and green). At the conserved segments, the homologous proteins display significant sequence conservation and structural similarity. } 
\label{ilpobjective}
\end{figure*}

The common framework identification problem can be described as:
given a collection of homologous proteins $H=\{s_1,\ldots,s_N\}$ with length $L_1,\ldots,L_N$, the objective is to find $m$ segments with length $n$ with high sequence conservation and structural similarity. As an example, Fig.~\ref{ilpobjective} shows the common frameworks shared by protein {\tt 3gxr\_A} and its homologous proteins.


\subsection{Basic idea of the linear program}

The common framework poses double-fold requirements, i.e., significantly high sequence conservation and structural similarity. In the linear program, the objective function was designed to describe structural similarity, and the constraints were designed to describe sequence similarities. 

Specifically, the linear program utilizes a set of boolean variables to represent the location of conserved segments, i.e., $x_{ij}^{k}=1$ denotes that in the $i$th protein, the $k$th segment is located at the $j$-th residue. Then, the structural similarity objective and sequence similarity constraints can be described using $x_{ij}^{k}$.

The constraints were designed to represent the following requirements.
\begin{itemize}
\item For any sequence, the $k$th segment in common framework is unique;
\item No segment in a common framework overlaps nor crosses. 
\item The segments should have significantly high sequence similarity. 
\end{itemize}
%
%
The integer linear programming model can be described as:\\
$$\ \ \ \ \ \ \ {\rm max}\ \ \ \ \ \ \ \  structural\ similarity\ \ \ \ \ \ \ \ \ \ \ \ \ \ \ \ \ \ \ \ \ \ \ \ \ \ \ \ \ \ \ \ \ \ \ \ \ \ \ \ \ \ \ \ \ \ \ \ \ \ \ \ \ \ \ \ \ \ \ \ \ \ \ \ \ \ \ \ \ \ \ \ \ $$
\begin{eqnarray}
{\rm s.t.}\ \ \ \ \ \ \ \ \ \ 			\mathop\sum\limits_{j=1}^{L_i}x_{ij}^{k} &=& 1,\ i=1,\ldots,N;k=1,\ldots,m \\
		        \mathop\sum\limits_{j=1}^{C}x_{ij}^{k_a}-\mathop\sum\limits_{j=1}^{C}x_{ij}^{k_b} &\geq& 0,\ i=1,\ldots,N;C=1,\ldots,L_i;1\leq k_a<k_b\leq m \\
				x_{ij_1}^{k}+x_{ij_2}^{k+1} &\leq& 1,\ i=1,\ldots,N;k=1,\ldots,m-1;1\leq j_2<j_1+n\leq L_i+n \\
				M_{i_1 j_1,i_2 j_2}*x_{i_1 j_1}^{k}*x_{i_2 j_2}^{k} &\geq& T,\ i_1,i_2=1,\ldots,N;j_1=1,\ldots,L_{i_1};j_2=1,\ldots,L_{i_2};k=1,\ldots,m
\end{eqnarray}
where $L_i$ denotes the length of the $i$th protein, and $M$ denotes the pre-calculated sequence similarity matrix. In particular, the cell $M_{i_1 j_1,i_2 j_2}$ denotes the sequence similarity between of the  segment starting from $j_1$ in the $i_1$th protein and the segment starting from $j_2$ in the $i_2$th protein.

\subsection{Refining the ILP model }

In our model, the structure similarity is described using Dscore~\cite{zhang2013fast}.
\begin{eqnarray}
Dscore(A,B) &=& \frac{1}{N^2}\mathop\sum\limits_{i=1}^{L}\mathop\sum\limits_{j=1}^{L} \frac{1}{1+(a_{ij}-b_{ij})^2} \nonumber\\
			  &\approx& \frac{1}{N^2}\mathop\sum\limits_{i=1}^{L}\mathop\sum\limits_{j=1}^{L} (1-(a_{ij}-b_{ij})^2) \nonumber
\end{eqnarray}
where $a_{ij}$ and $b_{ij}$ denote the C$_\alpha$ distance of residue $i$ and residue $j$ in protein $A$ and $B$ respectively.

The final integer linear programming can be formulated as:
$$\ \ \ \ \ \ \ \ \ \ \ \ \ \ \ \ \ \ \ \ \ \ \ \ \ \ \ \ \ \ \ \ \ \ \ \ \ \ \ \ \ {\rm max}\ \ \ \ \mathop\sum\limits_{i_1=1}^{N}\mathop\sum\limits_{i_2=1}^{i_1}\mathop\sum\limits_{j_{11}=1}^{L_{i_1}}\mathop\sum\limits_{j_{12}=1}^{j_{11}}\mathop\sum\limits_{j_{21}=1}^{L_{i_2}}\mathop\sum\limits_{j_{22}=1}^{j_{21}}\mathop\sum\limits_{k_1=1}^{m}\mathop\sum\limits_{k_2=1}^{k_1}D_{i_1 j_{11}j_{12},{i_2 j_{21}j_{22}}}^{k_1 k_2}*\tau_{i_1 j_{11}j_{12},{i_2 j_{21}j_{22}}}^{k_1 k_2}\ \ \ \ \ \ \ \ \ \ \ \ \ \ \ \ \ \ \ \ \ \ \ \ \ \ \ \ \ \ \ \ \ \ \ \ \ \ \ $$
\begin{eqnarray}
{\rm s.t.}\ \ \ \ \ \ \ \ \ \ \ \ \ \ x_{i_1 j_{11}}^{k_1}+x_{i_1 j_{12}}^{k_2}+x_{i_2 j_{21}}^{k_1}+x_{i_2 j_{22}}^{k_2}-\tau_{i_1 j_{11}j_{12},{i_2 j_{21}j_{22}}}^{k_1 k_2} &\leq& 3 \nonumber\\
		x_{i_1 j_{11}^{k_1}}-\tau_{i_1 j_{11}j_{12},{i_2 j_{21}j_{22}}}^{k_1 k_2} &\geq& 0 \nonumber\\
		x_{i_1 j_{12}^{k_2}}-\tau_{i_1 j_{11}j_{12},{i_2 j_{21}j_{22}}}^{k_1 k_2} &\geq& 0 \nonumber\\
		x_{i_2 j_{21}^{k_1}}-\tau_{i_1 j_{11}j_{12},{i_2 j_{21}j_{22}}}^{k_1 k_2} &\geq& 0 \nonumber\\
		x_{i_2 j_{22}^{k_2}}-\tau_{i_1 j_{11}j_{12},{i_2 j_{21}j_{22}}}^{k_1 k_2} &\geq& 0 \nonumber\\
		\mathop\sum\limits_{j=1}^{L_i}x_{ij}^{k} &=& 1 \nonumber\\
		\mathop\sum\limits_{j=1}^{C}x_{ij}^{k_a}-\mathop\sum\limits_{j=1}^{C}x_{ij}^{k_b} &\geq& 0 \nonumber\\
				x_{ij_1}^{k}+x_{ij_2}^{k+1} &\leq& 1 \nonumber\\
				x_{i_1 j_1}^{k}+x_{i_2 j_2}^{k}-t_{i_1 j_1,i_2 j_2}^{k} &\leq& 1 \nonumber\\
				x_{i_1 j_1}^{k}-t_{i_1 j_1,i_2 j_2}^{k} &\geq& 0 \nonumber\\
				x_{i_2 j_2}^{k}-t_{i_1 j_1,i_2 j_2}^{k} &\geq& 0 \nonumber\\
				M_{i_1 j_1,i_2 j_2}-t_{i_1 j_1,i_2 j_2}^{k}*T  &\geq& 0 \nonumber
\end{eqnarray}
where the indicator $\tau_{i_1 j_{11}j_{12},{i_2 j_{21}j_{22}}}^{k_1 k_2}$ equals to 1 iff all the four item $x_{i_1 j_{11}^{k_1}}$, $x_{i_1 j_{12}^{k_2}}$, $x_{i_2 j_{21}^{k_1}}$, $x_{i_2 j_{22}^{k_2}}$ equal to 1, and 0 otherwise.
The indicator $t_{i_1 j_1,i_2 j_2}^{k}$ equals to 1 iff both $x_{i_1 j_1}^{k}$ and $x_{i_2 j_2}^{k}$ equal to 1.
The $D$ and $M$ matrix are calculated in advance.
The cell $D_{i_1 j_{11}j_{12},{i_2 j_{21}j_{22}}}^{k_1 k_2}$ denotes the approximated Dscore of segments start from $j_{11}$ and $j_{21}$ in the $i_1$th protein and segments start from $j_{21}$ and $j_{22}$ in the $i_2$th protein.

\subsection{An example}

\begin{figure*}
\centering
\includegraphics[width=4in]{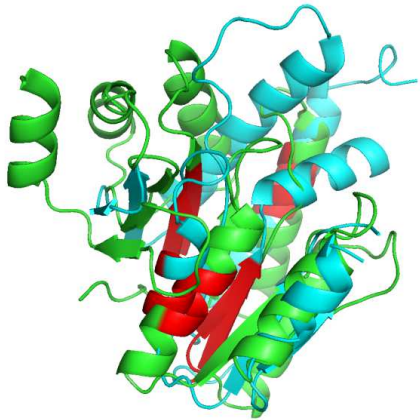}
\caption{Common frameworks shared by two domains in SCOP family c.37.1.11.}
\label{c37111}
\end{figure*}

Fig.~\ref{c37111} shows the common frameworks shared by two domains in SCOP family c.37.1.11.
The common frameworks has a Dscore of  8.35 and an RMSD of 1.9\AA, implying a significantly high structure conservation. 

\section{Experiments before CASP11}

For a total of over 27,000 proteins in PDB70, updated at Apr. 19, 2014, the common frameworks were
identified to yield a database called TOPO. The test set consists of 142 pairs of protein structures
similar in structure but with low sequence identity. Traditional threading approaches, say HHpred, fail
to build an accurate alignment between such protein pairs. In contrast, our alignment method
successfully build accruate alignment (TMscore$ > 0.4$) for seven protein pairs, and generate accurate
contact information for 45 protein pairs. Take a pair of protein 3dz1\_A vs. 1twd\_A as an example. The
two proteins share similar protein structure (TMscore=0.56); however, the alignment generated by
HHpred has a TMscore of only 0.22. In contrast, our alignment method generates an alignment with
TMscore=0.43.

\section{Conclusions}

Unlike close homology proteins, remote homology proteins show weakly overall sequence signals of
structure similarity. However, they still share common frameworks which carry strong sequence
signals of structure similarity. Aligning against the common frameworks instead of whole protein
sequences improves the fold recognition.

\section*{Acknowledgement}

The study was funded by the National Basic Research Program of China (973 Program) under Grant 2012CB316502, the National Nature Science Foundation of China under Grants 11175224 and 11121403, 31270834, 61272318, 30870572, and 61303161 and the Open Project Program of State Key Laboratory of Theoretical Physics (No.Y4KF171CJ1). This work made use of the eInfrastructure provided by the European Commission co-funded project CHAIN-REDS (GA no 306819).


\begin{thebibliography}{99}

\bibitem{zhang2013fast}

Zhang J, Xu D. Fast algorithm for population‐based protein structural model analysis[J]. \textit{Proteomics}, 2013, {\bf 13(2)}: 221-229.

\bibitem{TreeThreader}

Wu W, Chen G, Kan W, \textit{et al.} Harness public computing resources for protein structure prediction computing[C]. \textit{The International Symposium on Grids and Clouds (ISGC)}. 2013, 2013.

\bibitem{yang2008specific}

Yang Y, Zhou Y. Specific interactions for ab initio folding of protein terminal regions with secondary structures[J]. \textit{Proteins: Structure, Function, and Bioinformatics}, 2008, {\bf 72(2)}: 793-803.

\end{thebibliography}
\end{document}